\documentclass[reprint,aps,prb]{revtex4-2}

\usepackage[T1]{fontenc}
\usepackage{graphicx}
\usepackage{amsmath}
\usepackage{bm}

\begin{document}


\title{Nonlinear Optical Quantum Communication with a Two-Dimensional Perovskite Light Source} 



\author{Shuyue Feng}
\altaffiliation{Equal contributions.}
\affiliation{Department of Chemistry, University of North Carolina at Chapel Hill, Chapel Hill, NC 27599, USA}

\author{Zijian Gan}
\altaffiliation{Equal contributions.}
\affiliation{Department of Chemistry, University of North Carolina at Chapel Hill, Chapel Hill, NC 27599, USA}

\author{Camryn J. Gloor}
\affiliation{Department of Chemistry, University of North Carolina at Chapel Hill, Chapel Hill, NC 27599, USA}

\author{Wei You}
\affiliation{Department of Chemistry, University of North Carolina at Chapel Hill, Chapel Hill, NC 27599, USA}

\author{Andrew M. Moran}
\email{ammoran@unc.edu}
\affiliation{Department of Chemistry, University of North Carolina at Chapel Hill, Chapel Hill, NC 27599, USA}

\date{\today}


\date{\today}

\begin{abstract}
Two--dimensional organic-–inorganic hybrid perovskite (2D-OIHP) quantum wells are emerging as promising light sources for quantum communication technologies, owing to their ability to generate polarization--encoded optical signals. In this work, we explore how nonlinear optical phenomena can be exploited for quantum information applications, demonstrating the versatility that arises from resonant coupling among excited states. By tracking changes in the ellipticities of signal photons on femtosecond timescales in four--wave--mixing experiments, we first establish a method for information encoding based on exciton spin dynamics and biexciton correlations. Using single--photon detection, we then implement a proof-of-principle quantum communication protocol by mapping these polarization states onto binary sequences. While the polarizations of weak coherent pulses are typically manipulated with optical elements in traditional quantum key distribution approaches, the intrinsic electronic structure and spin relaxation processes within the 2D-OIHP system determine the characteristics of the signal photons in our method. As a demonstration, an ASCII message consisting of 56 bits is transmitted through the polarization states of photons emitted by 2D-OIHP quantum wells. These results show that the information transmission efficiency depends strongly on contributions from biexciton states, highlighting the potential of spin-dependent nonlinear optical processes for quantum communication.
\end{abstract}


\maketitle 

\section{Introduction}
Two--dimensional organic–inorganic hybrid perovskites (2D-OIHPs) are a versatile class of light-emitting materials that bridge classical optoelectronics and emerging quantum technologies \cite{yuanPerovskiteEnergyFunnels2016,smithTuningLuminescenceLayered2019,gaoMolecularEngineeringOrganic2019,leiEfficientEnergyFunneling2020,wangTwodimensionalHalidePerovskite2021,alvarado-leañosLasingTwoDimensionalTin2022,wangQuasi2DDionJacobsonPhase2023,guColorTunableLeadHalide2024,martinRemoteControlSteering2025}. Their quantum well structures support efficient light-emitting diodes and tunable lasers through strong excitonic confinement and high photoluminescence yields. Structural asymmetry introduced by chiral organic cations has recently enabled circularly polarized light emission, opening new opportunities for spin-selective devices \cite{maChiral2DPerovskites2019,liuBrightCircularlyPolarized2023,moroniChiral2DQuasi2D2024,liLargeExchangedrivenIntrinsic2024,liuDirectObservationCircularly2024,chenCircularlyPolarizedLight2019,hePerovskiteSpinLightemitting2025,liChiralQuasi2DPerovskites2025,dongChiralityInducedSpinSelectivity2025}. In addition, the demonstration of single-photon emission from colloidal perovskite quantum dots underscores the promise of this class of materials for integration into quantum photonic platforms \cite{utzatCoherentSinglephotonEmission2019,jagielskiScalablePhotonicSources2020,castellettoProspectsChallengesQuantum2022,farrowUltranarrowLineWidth2023,zhuSinglephotonSuperradianceIndividual2024,esmannSolidStateSinglePhotonSources2024}. While spin alignment can be induced by the absorption of circularly polarized light, subsequent relaxation typically occurs on picosecond or shorter timescales due to rapid population equilibration among excited states in 2D-OIHP materials composed of single lead-iodide layers \cite{toddDetectionRashbaSpin2019,taoDynamicPolaronicScreening2020,bourelleHowExcitonInteractions2020,chenTuningSpinPolarizedLifetime2021,songRolePolaronicStates2023,qinCoherentExcitonSpin2025,fengNonlinearOpticalSignatures2025,ganMotionalNarrowingSpin2025}. Although rapid spin depolarization limits the suitability of 2D-OIHPs for spintronic applications, it produces pronounced changes in the ellipticity of four-wave mixing signal fields, a distinctive optical signature that can be exploited for information encoding.

\begin{figure*}[!t]
\centering
\includegraphics[width=0.8\textwidth]{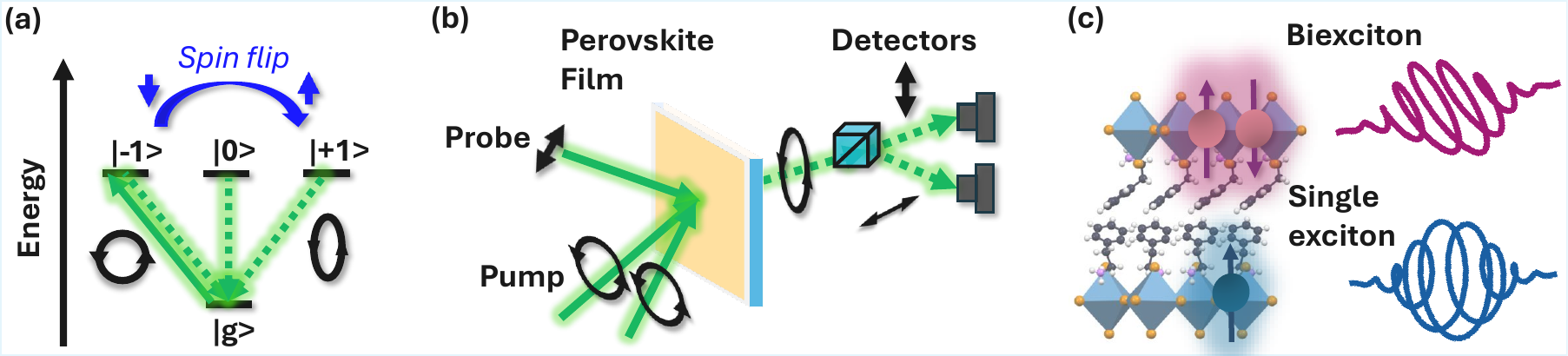}%
\caption{(a) Spin relaxation induces an elliptical-to-linear transformation in the polarization of the transient grating signal field. The relative amplitudes of the wave components that contribute to this effect are governed by the populations of the exciton states. (b) Circularly polarized light generates a macroscopic spin alignment within the ensemble of quantum wells. The radiated signal field is elliptically polarized due to the nonequilibrium distribution of populations within the exciton fine structure. (c) Single- and biexciton resonances radiate fields with distinct polarization ellipses, reflecting differences in spin dynamics and optical selection rules.}%
\label{figure1}
\end{figure*}

Here, we investigate the influence of spin relaxation on the polarization dynamics of four-wave mixing signals generated by a lead-iodide-based 2D-OIHP incorporating phenethylammonium (PEA) organic spacer layers. The (PEA)$_2$PbI$_4$ system exhibits three nearly degenerate exciton fine structure states that enable photoexcitation of nonequilibrium spin populations with circularly polarized light. Spin–orbit coupling among these exciton states, together with motional narrowing, subsequently eliminates the macroscopic spin polarization on the sub--100-fs timescale \cite{fengNonlinearOpticalSignatures2025,ganMotionalNarrowingSpin2025}. Using an experimental approach resembling ultrafast Faraday rotation spectroscopy \cite{baumbergFemtosecondFaradayRotation1994,odenthalSpinpolarizedExcitonQuantum2017,sutcliffeFemtosecondMagneticCircular2021,bourelleOpticalControlExciton2022,romanoCationTuningPolaron2025}, spin relaxation is first shown to induce an elliptical-to-linear transformation of the signal field polarization across the spectral range of single- and biexciton resonances in (PEA)$_2$PbI$_4$ (see Figure~\ref{figure1}). We then attenuate the signal pulses to the single-photon level and resolve their ellipticities using a photon-number-resolving detection method. The dynamics in the signal field ellipticities observed for the biexciton resonances are mapped onto binary sequences to implement the quantum communication experiment, requiring the accumulation of only tens of photons per bit \cite{shorSimpleProofSecurity2000,bennettQuantumCryptographyPublic2014}. Unlike traditional quantum communication approaches that manipulate the polarization states of weak coherent pulses using external optical elements, our method leverages the intrinsic electronic structure and spin-relaxation dynamics of the 2D-OIHP system to dynamically control the polarization basis of the light source. 

The key physical processes underlying this work are illustrated schematically in Figure~\ref{figure1}. Exciton states in 2D-OIHPs possess angular momentum fine structure, permitting macroscopic spin alignment upon excitation with circularly polarized light; however, spin-flip scattering among these states rapidly depolarizes the ensemble on sub--picosecond timescales. In this regime, nonlinear optical techniques are well-suited for tracking spin relaxation by monitoring changes in the polarization of the emitted signal field. Moreover, the polarization states of four--wave--mixing signal photons can be quantitatively modeled using an empirical response function for (PEA)$_2$PbI$_4$, established through conventional third-order nonlinear spectroscopies \cite{fengNonlinearOpticalSignatures2025,ganMotionalNarrowingSpin2025}. With the pulse configuration depicted in Figure~\ref{figure1}(b), the initial spin alignment gives rise to elliptically polarized four--wave--mixing signals, which progressively evolve toward linear polarizations as the exciton populations equilibrate. The ellipticity and waveform of the radiated field reflect the stochastic fluctuations within the material and vary markedly with the detection wavelength due to spectral overlap between single- and biexciton resonances. Among these transitions, biexcitons composed of antiparallel single-exciton spins generate signal fields with rotated polarization ellipses, an effect that arises uniquely from the nonlinear optical response. These higher-order optical effects are harnessed to minimize the number of photons required for accurate information transmission in the present quantum communication experiment.

\section{Experimental Methods}
\subsection{Materials Fabrication}

Phenethylammonium iodide (PEAI), dimethyl sulfoxide (DMSO), toluene, hydroiodic acid, and dimethylformamide (DMF) were sourced from Sigma-Aldrich and used without further purification. Lead iodide (99.9985\%) was obtained from Thermo Scientific. To fabricate (PEA)\textsubscript{2}PbI\textsubscript{4} thin films, PEAI and PbI\textsubscript{2} were dissolved in DMF at a 2:1 molar ratio, yielding a final Pb\textsuperscript{2+} concentration of 0.1~M. The resulting precursor solution was deposited on cleaned glass substrates using a spin-coating process at 5000~rpm for 30~s. 

Standard microscope slides (25 mm $\times$ 75 mm $\times$ 1.0 mm, Fisher Scientific) were sectioned into $\sim$25~mm~$\times$~25~mm squares for use as substrates. Substrates were cleaned by sequential sonication in deionized water, acetone, and isopropanol for 15~minutes each, then dried under a nitrogen stream and treated with UV--ozone for 15~minutes before film deposition.

All films were annealed at 100~$^\circ$C for 10~minutes immediately after spin coating. Spin coating was conducted entirely within a nitrogen-filled glovebox. Powder X-ray diffraction (PXRD) was performed under ambient conditions using a Rigaku SmartLab diffractometer with Cu~K$\alpha$ radiation in a parallel-beam configuration on intact, oriented films. Additional experimental details, including PXRD data, are available in our previous studies on (PEA)\textsubscript{2}PbI\textsubscript{4} \cite{ganElucidatingPhononDephasing2024,fengNonlinearOpticalSignatures2025}.

\subsection{Photon-Resolved Transient Grating Spectroscopy}
Four-wave mixing experiments were carried out using a Coherent Libra titanium–sapphire laser system that delivers 3.5-mJ, 50-fs pulses at a 1-kHz repetition rate. To generate a visible continuum beam, roughly 1.8~mJ of the 800~nm fundamental output was focused into a 4~m argon-filled tube maintained at an absolute pressure of 1.5--2.0~atm. The initial $\sim$5~mm beam diameter was reduced by a factor of 2.5 using a telescope composed of silver-coated spherical mirrors with focal lengths of 10~cm and 25~cm. The continuum was then routed through a fused-silica prism compressor, with the down-collimated laser beam diameter enabling a minimal spacing of 30~cm between the prism tips. The 490--550-nm spectral region was filtered using a slit placed in the spatially dispersed beam within the prism compressor. Further dispersion compensation was achieved using a chirped-mirror compressor with a total group-delay dispersion of $-880~\mathrm{fs^2}$ (Layertec). This approach resulted in an instrument response function with a temporal width of 20~fs at the sample position, consistent with sub-20-fs pulse durations.

\begin{figure}[ht]
    \centering
 \includegraphics[width=1.0\linewidth]{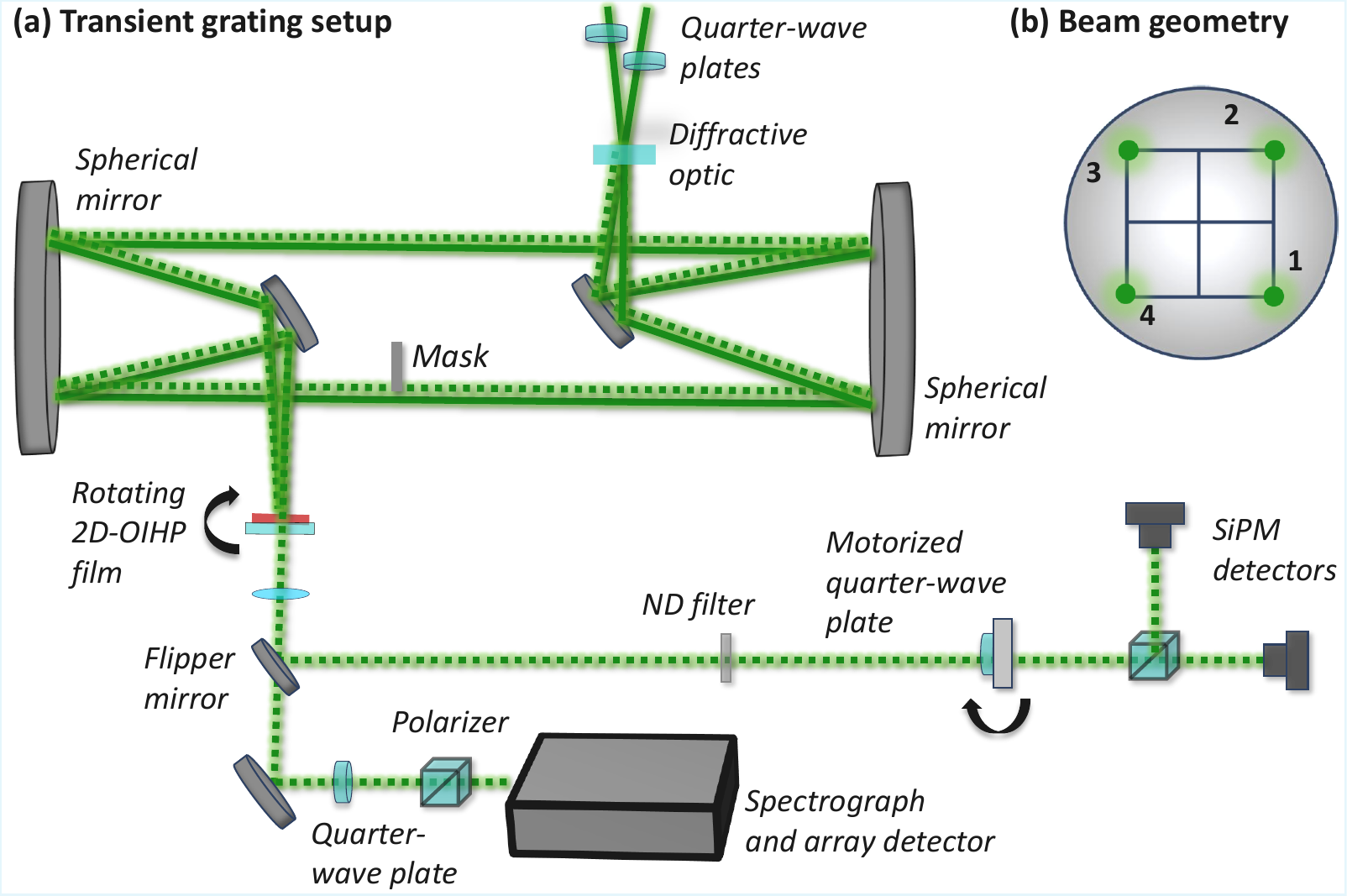}%
\caption{Transient-grating experiments were conducted with a diffractive-optic--based interferometer and square beam geometry. The initial pulse pair was circularly polarized, whereas the third pulse was vertically polarized. Conventional dispersed signal detection was accomplished with a spectrograph and array detector. Alternatively, the signal beam was attenuated to a mean fluence of $\sim$1~photon per pulse and detected with a pair of silicon photomultiplier modules (SiPM). The transient imbalance between horizontal and vertical signal polarizations was determined by passing the beam through the quarter-wave plate and polarizer after the sample. The quarter-wave plate in the photon-resolved detection path was mounted on a motorized rotation stage.}%
\label{figure2}
\end{figure}

As shown in Figure~\ref{figure2}, the laser beams were directed through 3.5-mm-thick wire-grid polarizers (Thorlabs WP25M-VIS) followed by 3.2-mm-thick achromatic quarter-wave plates (Thorlabs AQWP05M-580). The polarizations of the incoming pulses were preserved within the interferometer due to the custom diffractive optic (Holoeye), which provides equal diffraction efficiencies for horizontal and vertical polarization components \cite{fengNonlinearOpticalSignatures2025}. The two pump pulses were circularly polarized, and the probe pulse was vertically polarized for these experiments \cite{ganElucidatingPhononDephasing2024}. The diffractive optic produced an angular separation of 4.2$^\circ$ between the $\pm$1 diffraction orders at 520~nm, whereas this angle was reduced to 2.2$^\circ$ at the sample using a relay system composed of silver-coated spherical mirrors with focal lengths of 15 and 30~cm. After the sample, the signal beam passed through a quarter-wave plate and wire-grid polarizer to distinguish the horizontally and vertically polarized components. For conventional transient-grating signal detection, the light was directed into a home-built Czerny--Turner spectrograph consisting of two 20-cm focal-length, silver-coated mirrors and a 300-grooves/mm diffraction grating. Signal intensities were measured using a 12-bit monochrome CMOS camera (Thorlabs CS895MU) with a 100-ms integration time.

The signal beam was redirected using a flipper mirror for photon-resolved detection. The light was attenuated to mean values of 0.5--2.0~photons per pulse before passing through a quarter-wave plate mounted on a motorized rotation stage (Newport URB100CC). The orientation of the wave plate alternated between configurations in which the fast axis was either perpendicular to the vertically polarized third pulse in the transient-grating sequence or rotated by an angle of 45$^\circ$. The beam was then split into horizontally and vertically polarized components at a cube polarizer and detected using silicon photomultiplier (SiPM) modules with photon-resolving capabilities (Thorlabs PDA40). The two SiPM detectors yielded signals with 1.9-ns widths, which were simultaneously processed using an oscilloscope at a sampling interval of 2~ns (Siglent SDS1202X-E). The peaks of the signals were integer multiples of the single-photoelectron response, thereby permitting resolution of the number of photons arriving at each detector. Lens tubes containing a 500-nm bandpass filter with a 10-nm-wide transmission window were fastened to each of the SiPM detectors (Thorlabs FBH500-10). In addition, alignment disks with 2~mm apertures (Thorlabs SM1AP2) were placed at the entrances to the lens tubes to minimize contributions from stray light.

For all experiments, the laser pulses were focused to a 210-$\mu$m full-width-at-half-maximum spot with fluences of $\sim$10~$\mu$J/cm$^2$ at the sample position. The 2D-OIHP films were continuously spun at an angular frequency of 7.5~rad/s using a custom-built rotating mount to minimize photodegradation and spatially average the measured signals. In this setup, the laser spot traced a 1-cm-diameter circular path across the film. Under these conditions, no variations in signal intensities or spectral line shapes were observed during data acquisition periods of 1--2~hours \cite{ganElucidatingPhononDephasing2024}.

\section{Spin-Driven Polarization Effects in 2D-OIHP Systems}
\label{sec:background}

\begin{figure}[ht]
    \centering
 \includegraphics[width=0.55\linewidth]{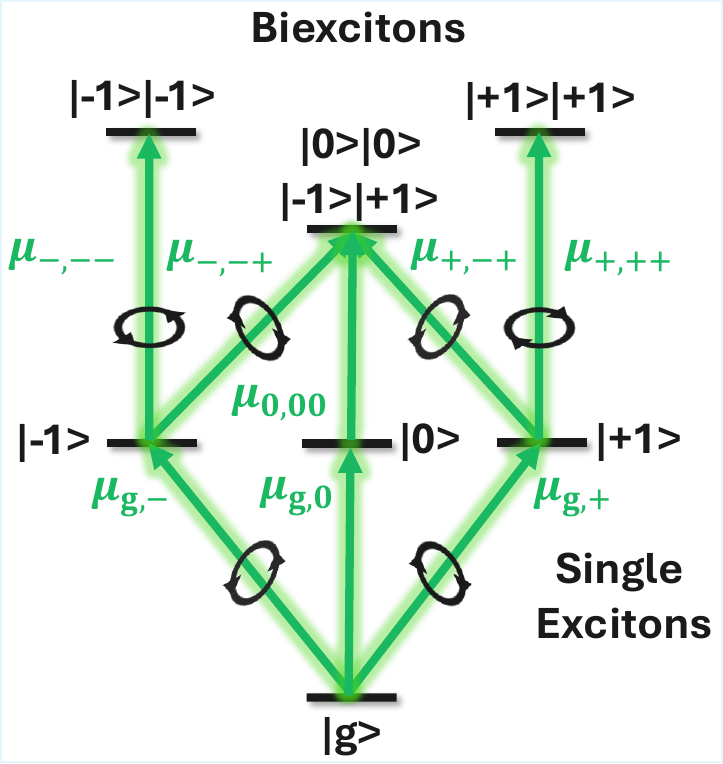}%
\caption{The (PEA)$_2$PbI$_4$ system exhibits a fine structure consisting of three bright single-exciton states with angular momentum quantum numbers $M_J = -1, 0, +1$. The basis set also includes three biexciton states with energies determined by whether the angular momentum vectors of the individual excitons have parallel or antiparallel orientations. The transition dipole moments, $\boldsymbol{\mu}_{g,+}$ and $\boldsymbol{\mu}_{g,-}$, are circularly polarized within the $x$–$y$ plane of the quantum well, whereas $\boldsymbol{\mu}_{g,0}$ is linearly polarized along the out-of-plane $z$-axis.}%
\label{figure3}
\end{figure}

While an empirical third-order response function for (PEA)$_2$PbI$_4$ has been developed in our recent work \cite{fengNonlinearOpticalSignatures2025,ganMotionalNarrowingSpin2025}, the behavior of the system is described here using a simplified model to clarify the polarization encoding mechanism utilized in our quantum communication procedure. As shown in Figure~\ref{figure3}, the band-edge electronic structure features a triplet of exciton fine-structure states, distinguished by their angular momentum projections onto the $z$-axis normal to the plane of the quantum well ($M_J = 0, \pm 1$). Because the energy splittings among these states are small compared to the spectroscopic linewidths \cite{sercelExcitonFineStructure2019,posmykQuantificationExcitonFine2022,posmykExcitonFineStructure2024}, they are treated as degenerate in our model of the system. Transient absorption measurements reveal resonances involving biexciton states at energies slightly above and below the single-exciton transition \cite{bourelleHowExcitonInteractions2020,fengNonlinearOpticalSignatures2025}. The polarization sensitivity of the optical response indicates that the higher- and lower-energy biexciton states correspond to configurations with parallel and antiparallel angular momentum vectors of the constituent excitons, respectively. Such alignment of the single-exciton angular momenta influences both the stabilities of the electronic states and the helicities of the optical transitions \cite{combescotTimeEvolutionTwo2010,yuEffectivemassModelMagnetooptical2016,beckerBrightTripletExcitons2018}.

Although the three single-exciton resonances share a common absorption line shape, a nonequilibrium population distribution within the exciton fine structure can be generated by selectively exciting the $M_J = \pm 1$ states with circularly polarized light of opposite handedness. In contrast, the transition dipole for the $M_J = 0$ exciton, which is linearly polarized along the out-of-plane $z$-axis, does not contribute to the macroscopic spin alignment. Transitions involving biexcitons with parallel and antiparallel spin are represented by a pair of independent Gaussian functions in our model, as further decomposition of the fine structure cannot be achieved due to line broadening at ambient temperatures. While the overall nonlinear optical response is captured using only three spectral line shapes, the underlying terms in the response function can still be resolved in four-wave mixing experiments by varying the polarizations of the incident laser pulses. Notably, the handedness of the transition dipoles depicted in Figure~\ref{figure3} remains an important factor in analyzing the optical response of solution-processed films, despite the softening of selection rules caused by the random orientations of quantum wells \cite{williamsImagingExcitedState2019,fengNonlinearOpticalSignatures2025}.

To illustrate the photon encoding mechanism employed below, the nonlinear response function is reduced to three classes of resonances with Gaussian line shapes, as shown in Figure~\ref{figure3}. The four-wave mixing signal is expressed as
\begin{equation}
S_{\alpha \beta \gamma \chi}\left(T, E_{\mathrm{det}}\right)
= \sum_{u=-1}^{1} B_{u,\alpha \beta \gamma \chi}\left(T\right)
\Phi_{u}\left(E_{\mathrm{det}}\right),
\label{eq:S_field}
\end{equation}
where $T$ is the population time and $E_{\mathrm{det}}$ is the energy of the dispersed signal spectrum.  The single-exciton resonance ($u=0$) represents the sum of the ground-state bleach and excited-state emission contributions, whereas the two excited-state absorption components correspond to the lower-energy ($u=-1$) and higher-energy ($u=1$) biexcitons. The complex line-shape function for resonance $u$ is given by
\begin{equation}
\Phi_{u}\left(E_{\mathrm{det}}\right)
= G_{u}\left(E_{\mathrm{det}}\right)
+ i\tilde{G}_{u}\left(E_{\mathrm{det}}\right),
\label{eq:Phi}
\end{equation}
where the absorptive part of the signal spectrum is defined as
\begin{equation}
G_{u}\left(E_{\mathrm{det}}\right)
= \exp\!\left[-\frac{\left(E_{\mathrm{det}} - u\Delta\right)^{2}}
{2\Delta^{2}}\right],
\label{eq:G_abs}
\end{equation}
and $\tilde{G}_{u}\!\left(E_{\mathrm{det}}\right)$ is the Hilbert transform of $G_{u}\!\left(E_{\mathrm{det}}\right)$.  The absorptive line shape is written using a single parameter $\Delta$ to capture both the line widths and spectral shifts of the biexcitons, as these quantities have similar magnitudes in (PEA)$_2$PbI$_4$. With this approach, the single-exciton resonance energy is located at the origin of the signal detection axis, while the two biexcitons appear at lower and higher energies. The relative signs and amplitudes of the resonances are incorporated into the real-valued expansion coefficients $B_{u,\alpha \beta \gamma \chi}\!\left(T\right)$, with the subscripts denoting the polarizations of the four laser pulses in increasing order from left to right (i.e., $\chi$ represents the emitted signal field).

While the detailed response function developed for (PEA)$_2$PbI$_4$ in prior work similarly assumed three distinct line shapes \cite{fengNonlinearOpticalSignatures2025,ganMotionalNarrowingSpin2025}, the present phenomenological approach offers a more streamlined and transparent treatment of polarization effects. Each term in the sum-over-states contained within the perturbative response function is weighted by the orientational average of the field–matter interactions, thereby incorporating the transition dipoles and polarizations of the electric fields. In contrast, the expression for the signal given in Equation~\eqref{eq:S_field} assigns each resonance $u$ an overall coefficient of $B_{u,\alpha \beta \gamma \chi}\!\left(T\right)$ rather than splitting the amplitude among multiple separately weighted contributions. This description is nonetheless sufficient to capture the nonlinear optical effect underlying our polarization encoding process, as the coefficients can be directly extracted from experimental data.

\begin{figure}[ht]
    \centering
 \includegraphics[width=1.0\linewidth]{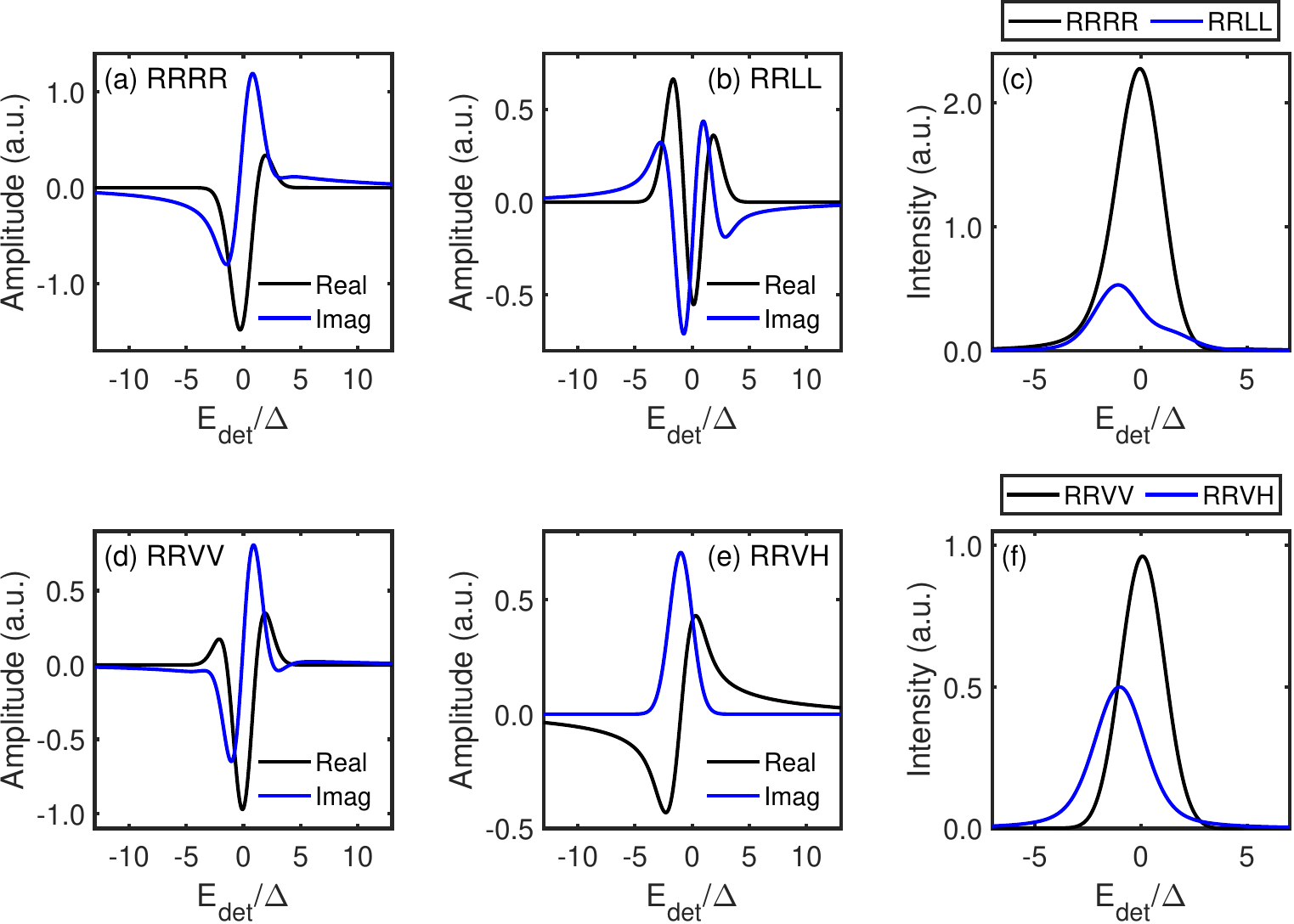}%
\caption{Four-wave mixing signal fields are calculated at a pump--probe delay of $T = 0$ using the parameters in Table~\ref{table1}. In this model, the single-exciton resonance energy is located at $E_{\mathrm{det}}=0$, whereas the biexcitons are spectrally shifted from the origin. Real (absorptive) and imaginary (dispersive) signal fields are shown for the (a) RRRR and (b) RRLL conditions, where R and L denote right- and left-handed circularly polarized fields, respectively. (c) The RRRR condition yields the largest signal intensity, while the peak of the RRLL spectrum is slightly red-shifted. Real and imaginary signal field components for the (d) RRVV and (e) RRVH conditions are also shown, where V and H denote vertical and horizontal linear polarizations, respectively. (f) The red shift of the RRVH signal intensity relative to RRVV reflects the linear combination of the RRRR and RRLL signals defined in Equation~\eqref{eq:signal_combination_VH}.
}%
\label{figure4}
\end{figure}

In Figures~\ref{figure4}(a)–\ref{figure4}(c), we present four-wave-mixing spectra calculated using the parameters summarized in Table~\ref{table1}, which are chosen to approximate the response of (PEA)$_2$PbI$_4$ \cite{fengNonlinearOpticalSignatures2025,ganMotionalNarrowingSpin2025}. Although the parameters do not represent a precise fit to experimental data, the targeted spectral signatures are robust to these differences and effectively capture the spin-dependent spectral features (PEA)$_2$PbI$_4$ and other 2D-OIHPs. Experimental configurations with the same- and opposite-circular pump and probe polarizations (SCP and OCP) are respectively denoted RRRR and RRLL, where R and L represent right- and left-handed circular polarizations. Notably, the handedness of the fields is arbitrary, as the LLLL and LLRR conditions yield physically indistinguishable results. The primary difference between the SCP and OCP spectra is the enhanced contribution from the lower-energy biexciton resonance in the OCP configuration. This produces a Gaussian peak near $E_{\mathrm{det}} = - \Delta$ in the real (absorptive) part of the signal field, whereas this resonance is negligible in the SCP spectrum (see Figures~\ref{figure4}(a) and~\ref{figure4}(b)).

Nonequilibrium population distributions within the exciton fine structure give rise to ellipticity of the four-wave mixing signal field when 2D-OIHPs are probed with linearly polarized light. Because the optical response reflects isotropic quantum-well orientations \cite{fengNonlinearOpticalSignatures2025,williamsImagingExcitedState2019}, the tensor components are described by the following linear combinations \cite{andrewsThreedimensionalRotationalAverages1977}:
\begin{equation}
\begin{split}
S_{RRVV}\!\left(T, E_{\mathrm{det}}\right)
&= \frac{1}{2} S_{RRRR}\!\left(T, E_{\mathrm{det}}\right) \\
&\quad + \frac{1}{2} S_{RRLL}\!\left(T, E_{\mathrm{det}}\right),
\end{split}
\label{eq:signal_combination_VV}
\end{equation}
and
\begin{equation}
\begin{split}
S_{RRVH}\!\left(T, E_{\mathrm{det}}\right)
&= -\frac{i}{2} S_{RRRR}\!\left(T, E_{\mathrm{det}}\right) \\
&+ \frac{i}{2} S_{RRLL}\!\left(T, E_{\mathrm{det}}\right).
\end{split}
\label{eq:signal_combination_VH}
\end{equation}
Here, the probe pulse is vertically-polarized in the lab frame (third subscript corresponding to V), whereas the signal pulse is decomposed into horizontal and vertical polarizations (fourth subscript V or H). The initial ellipticity of the signal field near time zero $T=0$ reflects the phase shift of the horizontal polarization component induced by the imaginary coefficients in Equation~\eqref{eq:signal_combination_VH}. However, the signal becomes vertically polarized within hundreds of femtoseconds because $S_{RRVH}(T, E_{\mathrm{det}})$ converges to zero as population equilibration drives a cancellation between the RRRR and RRLL tensor elements.  To further illustrate how spin relaxation influences the signal under the RRVH condition, Equations~\eqref{eq:S_field} and~\eqref{eq:signal_combination_VH} are combined to obtain
\begin{equation}
\begin{split}
&S_{RRVH}\!\left(T, E_{\mathrm{det}}\right) \\
&= -\frac{i}{2} \sum_{u=-1}^{1}
\!\left\{ B_{u,RRRR}\!\left(T\right)
- B_{u,RRLL}\!\left(T\right) \right\}
\Phi_{u}\!\left(E_{\mathrm{det}}\right).
\end{split}
\label{eq:rrvh_sum}
\end{equation}
Assuming an exponential spin-flip rate constant $k_{\mathrm{spin}}$ , the decay in the horizontally polarized signal component can be approximated as
\begin{equation}
\begin{split}
&S_{RRVH}\!\left(T, E_{\mathrm{det}}\right) \approx -\frac{i}{2} \exp(-k_{\mathrm{spin}} T)\\
&\times 
\sum_{u=-1}^{1}
\!\left\{ B_{u,RRRR}\!\left(0\right)
- B_{u,RRLL}\!\left(0\right) \right\}
\Phi_{u}\!\left(E_{\mathrm{det}}\right).
\end{split}
\label{eq:rrvh_decay}
\end{equation}
Because the two terms in $S_{RRVV}$ constructively interfere, the signal field becomes vertically polarized on the timescale of population equilibration within the exciton fine structure, $k_{\mathrm{spin}}^{-1}$, consistent with our model based on the full response function \cite{fengNonlinearOpticalSignatures2025,ganMotionalNarrowingSpin2025}.

Our polarization encoding mechanism leverages the interference effects described by Equations~\eqref{eq:signal_combination_VV}-\eqref{eq:signal_combination_VH}. The RRVV and RRVH signal fields displayed in Figures~\ref{figure4}(d) and~\ref{figure4}(e) exhibit differences in their line shapes arising from the cancellation of terms in Equation~\eqref{eq:signal_combination_VH}. While all three resonances are apparent under the RRVV polarization condition, the lowest-energy biexciton dominates the response with RRVH polarizations. Moreover, the dispersive part of the RRVH line shape extends to energies far below resonance, reflecting minimal interference with other transitions. The overall four-wave mixing signal intensities plotted in Figure~\ref{figure4}(f) illustrate the resulting red shift in the peak of the RRVH signal predicted at $T=0$. For (PEA)$_2$PbI$_4$ and other 2D-OIHPs examined by our group \cite{fengNonlinearOpticalSignatures2025,ganMotionalNarrowingSpin2025}, the time-zero RRVH signal intensity exceeds the RRVV intensity for detection energies below the biexciton resonance, whereas the signal field becomes entirely vertically polarized at delay times $T>> k^{-1}_{spin}$. This distinctive behavior is central to the transmission of quantum information in the present work.

\begin{table}[ht]
\centering
\caption{Expansion coefficients for spectroscopic model.}
\begin{tabular}{|c|c|c|}
\hline
\textbf{Coefficient} & \textbf{RRRR} & \textbf{RRLL} \\
\hline
$B_{-1,\alpha \beta \gamma \chi}(0)$ & 0 & $\sqrt{2}$ \\
\hline
$B_{0,\alpha \beta \gamma \chi}(0)$ & $-2$ & $-1$ \\
\hline
$B_{+1,\alpha \beta \gamma \chi}(0)$ & 1 & 1 \\
\hline
\end{tabular}
\vspace{2mm}
\label{table1}
\end{table}

\section{Experimental Results and Discussion}

While absorption of circularly polarized light induces a macroscopic spin polarization in 2D-OIHPs, the observed signatures of spin relaxation depend on the polarization state of the subsequent probe pulse. Using an approach analogous to ultrafast Faraday rotation spectroscopy \cite{baumbergFemtosecondFaradayRotation1994,odenthalSpinpolarizedExcitonQuantum2017,sutcliffeFemtosecondMagneticCircular2021,bourelleOpticalControlExciton2022,romanoCationTuningPolaron2025}, we find that the four-wave mixing signal fields generated by 2D-OIHPs such as (PEA)$_2$PbI$_4$ undergo a rapid elliptical-to-linear polarization transformation due to population equilibration among the exciton fine-structure states \cite{fengNonlinearOpticalSignatures2025}. We first investigate transient grating signal intensities in a configuration that includes a quarter-wave plate positioned after the sample, enabling flexible transformation of the polarization basis at the detectors. Knowledge of the system’s behavior is subsequently applied to optimize the encoding and transmission of information.

\subsection{Four-Wave Mixing Polarization Effects with Conventional Signal Detection}
\label{sec:convdet}

The present quantum communication experiment is implemented by inserting a quarter-wave plate with variable orientation between the sample and polarizer, as illustrated in Figure~\ref{figure2}. Although this introduces complexity beyond the standard orientational averaging of transition dipoles in four-wave mixing, the effect of the quarter-wave plate on the detected signal can be accurately modeled and serves as a practical means for manipulating the helicity of the signal field. We perform transient grating experiments with conventional signal detection to establish how the signal field depends on the pump-probe delay, emission wavelength, and quarter-wave plate orientation. These results are then used to define the optimal conditions for quantum communication.

\begin{figure}[ht]
    \centering
 \includegraphics[width=1.0\linewidth]{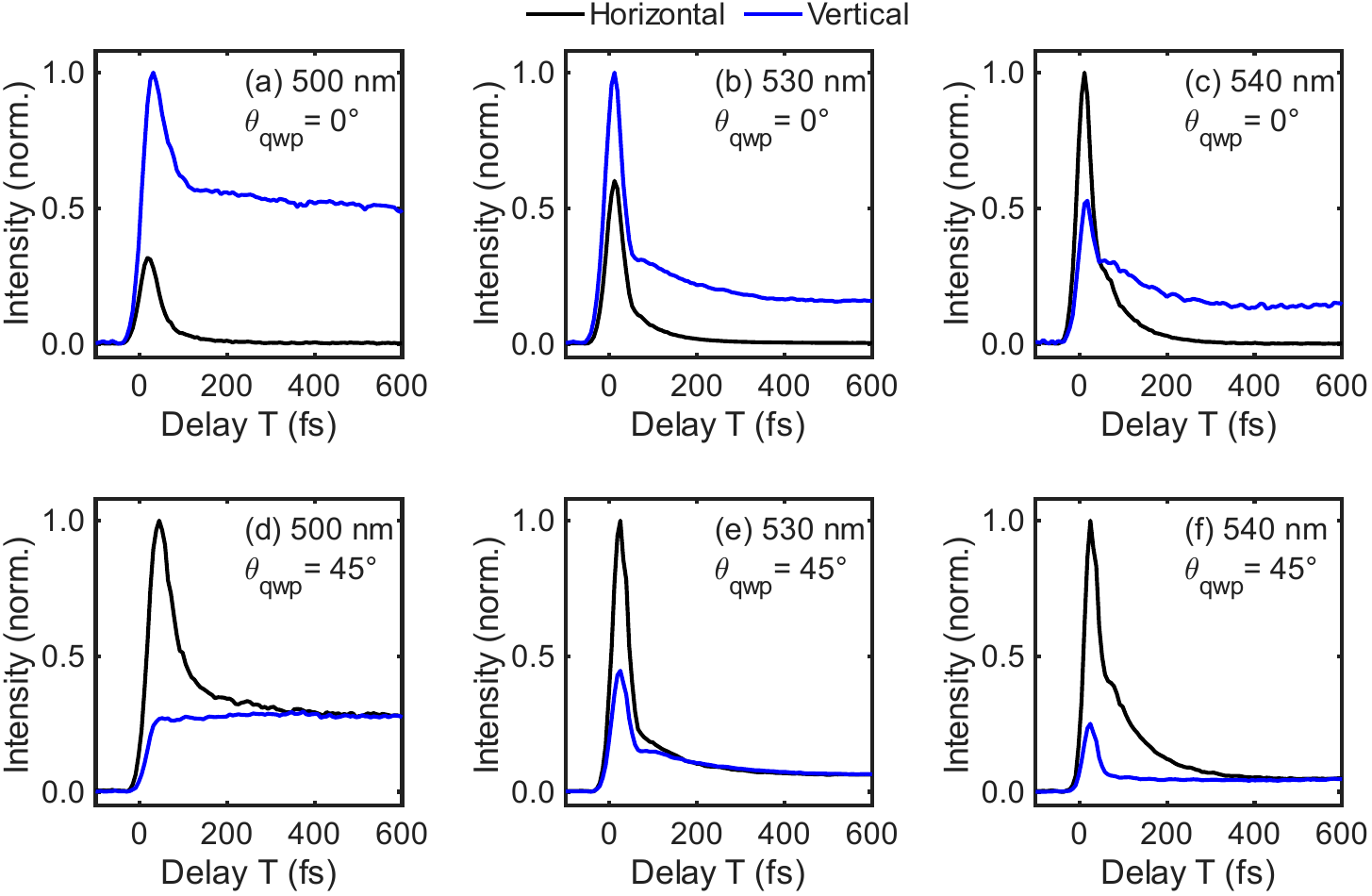}%
\caption{%
Transient grating signal intensities measured at the indicated quarter-wave plate angles and detection wavelengths. 
The angle between the vertically polarized probe and the fast axis of the quarter-wave plate used for signal analysis is set to either 
(a)--(c)~$\theta_{\mathrm{qwp}} = 0^{\circ}$ or (d)--(f)~$\theta_{\mathrm{qwp}} = 45^{\circ}$. 
The relative intensities of the horizontally and vertically polarized signals vary weakly across the 500--525~nm range but show significant dispersion in the spectral region corresponding to the lower-energy biexciton. 
In each panel, the intensities are normalized to the peak value of the signal with the largest magnitude.%
}
\label{figure5}
\end{figure}

To confirm the behaviors predicted in Section~\ref{sec:background} and demonstrate the feasibility of the quantum communication approach, four--wave mixing signals are measured under four polarization conditions using interleaved, back--to--back acquisitions. The polarization contrast for the signal intensity is given by
\begin{equation}
P\!\left( T, \lambda_{\mathrm{det}}, \theta_{\mathrm{qwp}} \right)
= 
\frac{I_{H}\!\left( T, \lambda_{\mathrm{det}}, \theta_{\mathrm{qwp}} \right)
 - I_{V}\!\left( T, \lambda_{\mathrm{det}}, \theta_{\mathrm{qwp}} \right)}
{I_{H}\!\left( T, \lambda_{\mathrm{det}}, \theta_{\mathrm{qwp}} \right)
 + I_{V}\!\left( T, \lambda_{\mathrm{det}}, \theta_{\mathrm{qwp}} \right)},
\label{eq:polarization_contrast}
\end{equation}
where \(T\) is the delay between the pump pulse-pair and probe, 
\(\lambda_{\mathrm{det}}\) is the detection wavelength, 
and \(\theta_{\mathrm{qwp}}\) is the angle between the fast axis of the quarter-wave plate and the vertically polarized probe field. 
In Figures~\ref{figure5}(a)--\ref{figure5}(c), signals acquired with \(\theta_{\mathrm{qwp}} = 0^{\circ}\) reveal a dominant vertically polarized signal component across the 500–530~nm spectral range; however, the horizontally polarized contribution rises sharply and surpasses the vertical intensity near 540~nm, as illustrated in Figure~4(f). 
When the quarter-wave plate is set to \(\theta_{\mathrm{qwp}} = 45^{\circ}\), the horizontally polarized signal similarly increases with detection wavelength but remains weaker than the vertical intensity near time zero (see Figures~\ref{figure5}(d)--\ref{figure5}(f)).

To reconstruct the signal field from polarization-resolved intensity measurements, we employ a numerical optimization approach based on a discrete search over the field amplitudes and phase differences between the horizontal and vertical polarization components. The electric field radiated by the 2D-OIHP system is written as \cite{fengNonlinearOpticalSignatures2025}: 
\begin{equation}
\vec{E}_{S}\!\left(T, \lambda_{\mathrm{det}}, t \right)
=   \exp\!\left(-i\,\frac{2\pi c}{\lambda_{\mathrm{det}}}\, t\right)
{{\vec{\varepsilon }}_{S}}\left( T,\lambda _{\mathrm{det}} \right)
\label{eq:ES}
\end{equation}
where
\begin{equation}
\begin{split}
{{\vec{\varepsilon }}_{S}}\left( T,\lambda _{\mathrm{det}}\right)
=&{{A}_{H}}\left( T,\lambda_{\mathrm{det}} \right)
\exp \left[ i\varphi \left( T,\lambda_{\mathrm{det}} \right) \right]
{{\hat{e}}_{H}} \\
&+{{A}_{V}}\left( T,\lambda _{\mathrm{det}} \right)
{{\hat{e}}_{V}}.
\end{split}
\label{eq:epsS}
\end{equation}
Here, $A_{H}\!\left(T, \lambda_{\mathrm{det}} \right)$ and $A_{V}\!\left(T, \lambda_{\mathrm{det}} \right)$ denote the horizontally and vertically polarized wave amplitudes, and $\varphi\!\left(T, \lambda_{\mathrm{det}} \right)$ is the spectral phase difference.  The solution is obtained through a numerical optimization procedure that jointly adjusts $A_{H}\!\left(T, \lambda_{\mathrm{det}} \right)$ and $\varphi\!\left(T, \lambda_{\mathrm{det}} \right)$.  

The detection process is modeled by propagating the field through a quarter-wave plate followed by a linear polarizer oriented in either the horizontal or vertical direction. The transformation matrix for a given fast-axis angle ${{\theta }_{\mathrm{qwp}}}$ is written as
\begin{equation}
\mathbf{Q}({{\theta }_{\mathrm{qwp}}})=\mathbf{\Theta }(-{{\theta }_{\mathrm{qwp}}})
\begin{pmatrix}
1 & 0  \\
0 & i  \\
\end{pmatrix}
\mathbf{\Theta }({{\theta }_{\mathrm{qwp}}}),
\end{equation}
where $\mathbf{\Theta }({{\theta }_{\mathrm{qwp}}})$ is the standard 2D rotation matrix:
\begin{equation}
\mathbf{\Theta }\left( {{\theta }_{\mathrm{qwp}}} \right)=
\begin{pmatrix}
\cos \left( {{\theta }_{\mathrm{qwp}}} \right) & -\sin \left( {{\theta }_{\mathrm{qwp}}} \right)  \\
\sin \left( {{\theta }_{\mathrm{qwp}}} \right) & \cos \left( {{\theta }_{\mathrm{qwp}}} \right)  \\
\end{pmatrix}.
\end{equation}
This transformation is evaluated in two orientations: ${{\theta }_{\mathrm{qwp}}}=0^{\circ}$ and $45^{\circ}$. For each configuration, we compute the field after the quarter-wave plate and project it onto the horizontal and vertical polarization axes using
\begin{equation}
\mathbf{\Pi}_{H} =
\begin{pmatrix}
1 & 0 \\
0 & 0
\end{pmatrix}
\end{equation}
and
\begin{equation}
\mathbf{\Pi}_{V} =
\begin{pmatrix}
0 & 0 \\
0 & 1
\end{pmatrix}.
\end{equation}
The resulting signal intensities are
\begin{equation}
{{I}_{H}}\left( T,\lambda_{\mathrm{det}}^{{}},{{\theta }_{\mathrm{qwp}}} \right)
={{\left| {{\mathbf{\Pi}}_{H}}\cdot \mathbf{Q}\left( {{\theta }_{\mathrm{qwp}}} \right)\cdot {{{\vec{\varepsilon }}}_{S}}\left( T,\lambda_{\mathrm{det}}^{{}} \right) \right|}^{2}}
\end{equation}
and
\begin{equation}
{{I}_{V}}\left( T,\lambda _{\mathrm{det}}^{{}},{{\theta }_{\mathrm{qwp}}} \right)
={{\left| \mathbf{\Pi}_{V}^{{}}\cdot \mathbf{Q}\left( {{\theta }_{\mathrm{qwp}}} \right)\cdot {{{\vec{\varepsilon }}}_{S}}\left( T,\lambda _{\mathrm{det}}^{{}} \right) \right|}^{2}}.
\end{equation}

These expressions are evaluated on a two-dimensional grid with the phase spanning $-\pi/2$ to $\pi/2$ in 0.01-radian steps, and the total intensity of the field normalized to~1:
\begin{equation}
A_{H}^{2}\left( T,\lambda_{\mathrm{det}}^{{}} \right)+A_{V}^{2}\left( T,\lambda_{\mathrm{det}}^{{}} \right)=1.
\end{equation}
The intensity ratio between horizontally and vertically polarized signal intensities is given by 
\begin{equation}
\Gamma\left( T,\lambda_{\mathrm{det}},{{\theta }_{\mathrm{qwp}}} \right)
=\frac{{{I}_{H}}\left( T,\lambda_{\mathrm{det}},{{\theta }_{\mathrm{qwp}}} \right)}{{{I}_{V}}\left( T,\lambda_{\mathrm{det}},{{\theta }_{\mathrm{qwp}}} \right)+\xi },
\end{equation}
where $\xi$ is a small regularization constant to avoid division by zero. At each point on the grid, the simulated ${{\Gamma}_{\mathrm{sim}}}\left( T,\lambda_{\mathrm{det}},{{\theta }_{\mathrm{qwp}}} \right)$ and measured ${{\Gamma}_{\mathrm{meas}}}\left( T,\lambda_{\mathrm{det}},{{\theta }_{\mathrm{qwp}}} \right)$ ratios in signal intensities are combined to define a total squared error:
\begin{equation}
\begin{split}
&SE\left( A_{H}^{{}},\varphi  \right) \\
&=\sum_{\substack{
   \theta_{\mathrm{qwp}} \\
   = \left[ 0^{\circ}, 45^{\circ} \right]
}}
{{{\left[ {{\Gamma}_{\mathrm{sim}}}\left( T,\lambda_{\mathrm{det}},{{\theta }_{\mathrm{qwp}}} \right)
-{{\Gamma}_{\mathrm{meas}}}\left( T,\lambda_{det},{{\theta }_{\mathrm{qwp}}} \right) \right]}^{2}}}.
\end{split}
\label{eq:SE}
\end{equation}
The optimal values of $A_{H}\left( T,\lambda_{\mathrm{det}} \right)$ and $\varphi \left( T,\lambda_{\mathrm{det}} \right)$ are those that minimize this error. The vertical amplitude is then inferred from normalization, such that the total intensity equals unity at each delay point.

\begin{figure}[ht]
    \centering
 \includegraphics[width=1.0\linewidth]{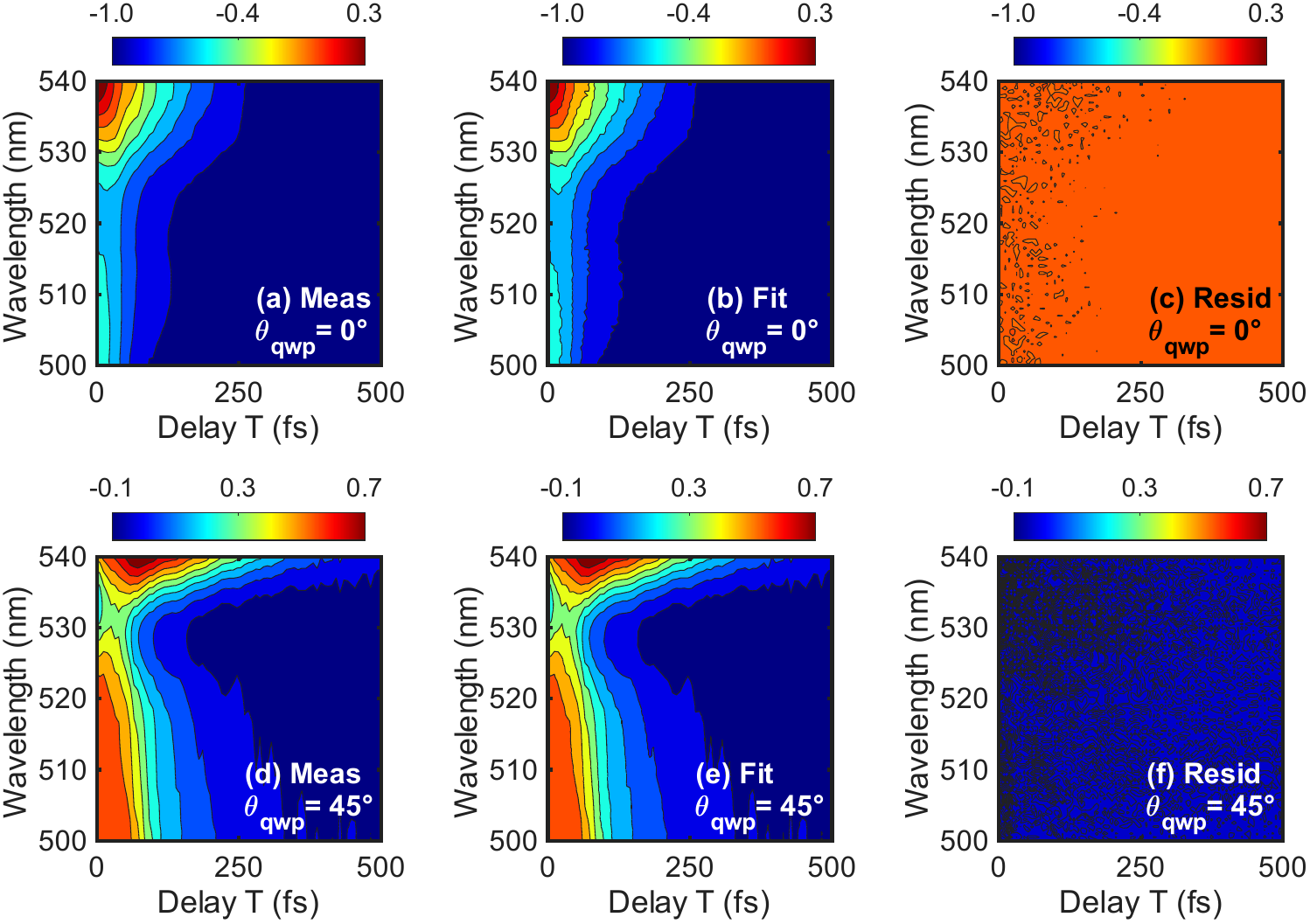}%
\caption{%
Measured and fitted polarization contrasts $P\!\left( T, \lambda_{\mathrm{det}}, \theta_{\mathrm{qwp}} \right)$ are compared at quarter-wave plate angles of 
(a)--(b)~$\theta_{\mathrm{qwp}} = 0^{\circ}$ and (d)--(e)~$\theta_{\mathrm{qwp}} = 45^{\circ}$. 
Residuals in the fitted signals are less than 5\% over the full range of pulse delays and detection wavelengths, as shown in panels (c) and (f). 
The fitted signals are generated using the numerical optimization routine described in Equations \eqref{eq:ES}--\eqref{eq:SE}.%
}
\label{figure6}
\end{figure}

Measured and fitted polarization contrasts are presented in Figure~\ref{figure6} to demonstrate the accuracy of the numerical optimization procedure. 
For both quarter-wave plate angles, the residual deviations are less than 5\% across the full range of pulse delays and detection wavelengths. 
These contour plots complement the temporal profiles of signal components presented in Figure~\ref{figure5} by revealing broader trends in both dimensions. 
Most notably, distinct behaviors are observed at detection wavelengths above 530~nm, where the resonance associated with the lower-energy biexciton is located. 
The simplified model introduced in Section~\ref{sec:background} clarifies that the horizontally polarized signal intensity dominates in this spectral region near time zero due to interferences within the nonlinear response function; however, the signal field becomes entirely vertically polarized on the $\sim$100-fs timescale of spin relaxation. 
Although the presence of the quarter-wave plate after the sample partially obscures these dynamics in the raw data, they are successfully captured in the reconstructed signal field.

\begin{figure}[ht]
    \centering
 \includegraphics[width=1.0\linewidth]{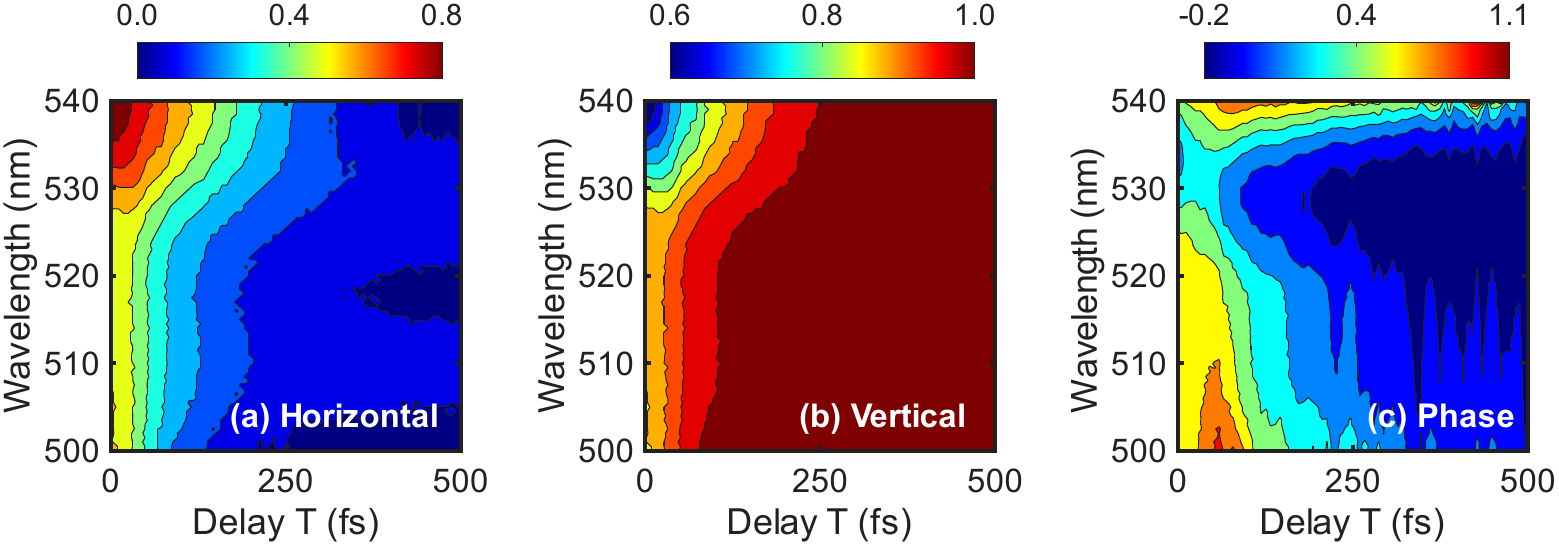}%
\caption{%
Parameters of the reconstructed electric field ${{\vec{\varepsilon }}_{S}}\left( T,\lambda _{\mathrm{det}}\right)$ include the (a) horizontally ($A_{H}$) and (b) vertically ($A_{V}$) polarized wave amplitudes. (c) The spectral phase shows that the signal field maintains consistent helicity across the entire spectrum, despite contributions from transition dipoles with opposite handedness. 
The total intensity, $A_{H}^{2} + A_{V}^{2}$, is normalized to unity.%
}
\label{figure7}
\end{figure}

In Figure~\ref{figure7}, we present the three parameters of the reconstructed signal field defined in Equations~\eqref{eq:ES}--\eqref{eq:epsS}. 
For each pump--probe delay time, the horizontally and vertically polarized amplitudes remain relatively uniform within the 500--530~nm wavelength range, while the phase difference between the two components reflects the ellipticity of the signal field. 
The horizontal amplitude fully vanishes within $\sim$200~fs due to spin relaxation, and the signal field becomes vertically polarized, as discussed in Section~\ref{sec:background}. 
At time zero, the horizontal and vertical polarization components have comparable magnitudes in the 530--540-nm wavelength range, with $A_{H}$ exceeding $A_{V}$ near 540~nm. 
The peak of the lower-energy biexciton resonance is located near 530~nm, whereas 540~nm corresponds to the long-wavelength ``tail'' of the spectral line shape (i.e., the enhanced region for the RRVH tensor element). 
Although the higher- and lower-energy biexcitons of (PEA)$_2$PbI$_4$ correspond to circularly polarized transition dipoles with opposite handedness, they radiate signal fields with the same helicity when probed with linearly polarized light. 
This behavior is reflected by the consistent sign of the phase difference across the full wavelength range near time zero in Figure~\ref{figure7}. 
Rather than originating from helicity inversion, the prominent horizontal polarization amplitude observed near 540~nm arises from a spectral shift in the signal field for the RRVH tensor element, as illustrated in Figure~\ref{figure4}(f).

\subsection{Quantum Communication by Transient Grating Polarization Contrasts}
The BB84 quantum key distribution protocol is one of the foundational methods in quantum information science. It was developed by Charles Bennett and Gilles Brassard in 1984 to address the problem of securely distributing cryptographic keys over a potentially insecure communication channel \cite{bennettQuantumCryptographyPublic2014,shorSimpleProofSecurity2000}. When implemented using the polarization states of individual photons, both the sender (traditionally called Alice) and the receiver (Bob) independently and randomly choose between two polarization bases, commonly the rectilinear (horizontal/vertical) and diagonal ($+45^{\circ}/-45^{\circ}$) bases, using optical elements such as wave plates. After transmission, Alice and Bob use a classical communication channel to compare the bases used for each photon, without revealing the actual bit values. They discard the cases where their bases do not match. In this scheme, each matching basis corresponds to one of four possible measurement conditions, with two of those mapped to binary 0 and the other two to binary 1. The resulting sifted key is provably secure against eavesdropping, provided that the light is transmitted at the single-photon level, thereby preventing an eavesdropper from gaining information without disturbing the quantum states in a detectable way.

\begin{figure*}[!t]
\centering
\includegraphics[width=0.85\textwidth]{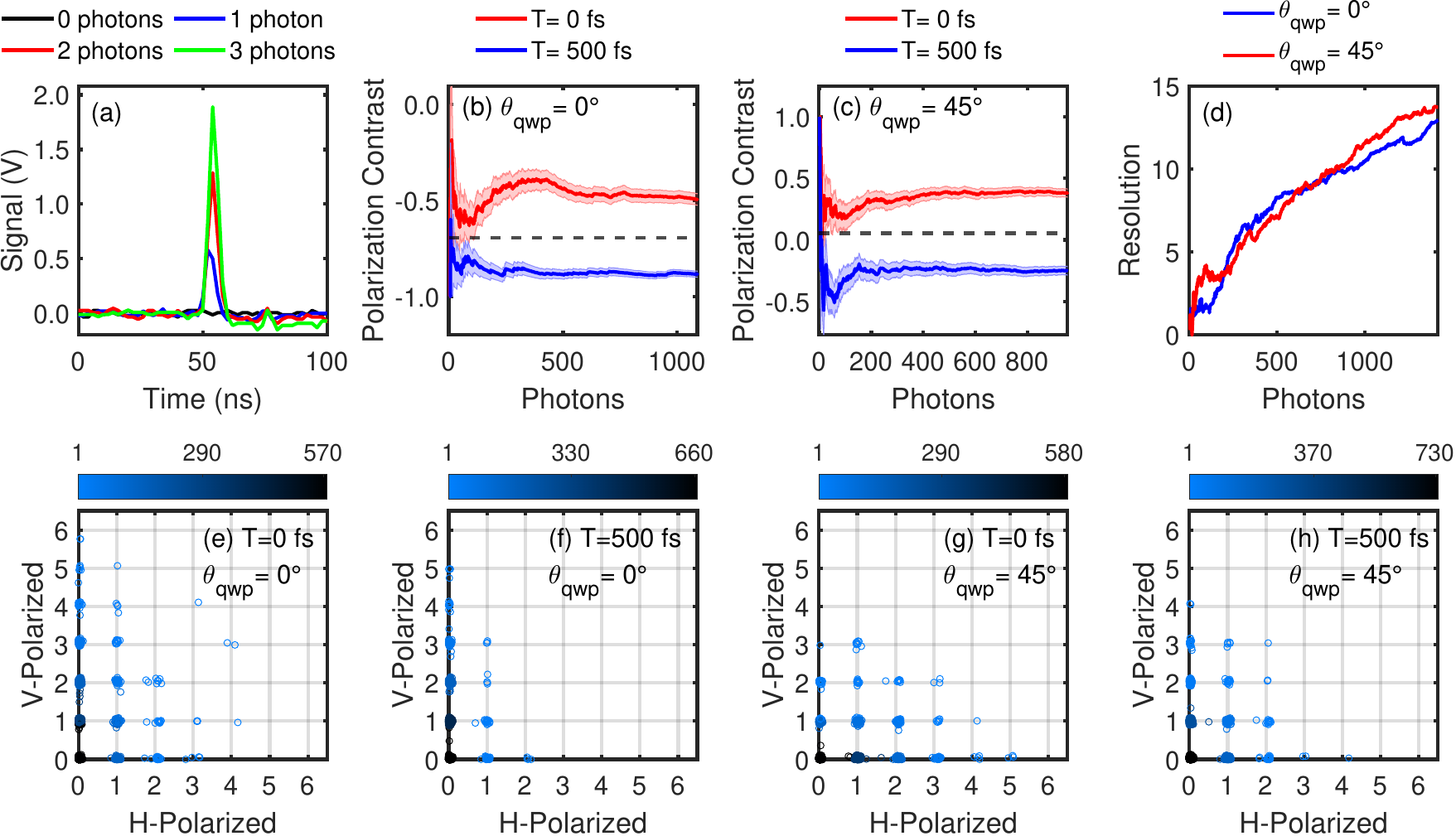}
\caption{Photon-resolved transient grating signals measured at a detection wavelength of 500~nm. (a) Signals acquired with the silicon photomultiplier detectors represent integer multiples of the voltage corresponding to a single photoelectron. (b)–(c) Convergence of the mean values $\bar{P}^{(M)}$ and standard errors $\sigma_{P}^{(M)}$ for polarization contrasts measured at the indicated pump–probe delay times and quarter-wave plate angles $\theta_{\mathrm{qwp}}$. (d) Resolution of the polarization contrasts $R_\mathrm{res}^{(M)}$ acquired for quarter-wave plate angles of $\theta_{\mathrm{qwp}} = 0^{\circ}$ and $45^{\circ}$. (e)–(h) Total numbers of horizontally and vertically polarized photons accumulated at the indicated pump–probe delay times and quarter-wave plate angles.}
\label{figure8}
\end{figure*}

In our approach, the four-wave mixing signal field generated by (PEA)$_2$PbI$_4$ is attenuated to an average of approximately one photon per pulse for use as the light source. The polarization basis is selected after the sample by randomly toggling the quarter-wave plate angle $\theta_{\mathrm{qwp}}$ between $0^{\circ}$ and $45^{\circ}$, as demonstrated in Section~\ref{sec:convdet}. In addition, two pump--probe delay times are randomized to produce either elliptically or linearly polarized signal photons at $T=0$ and $500~\mathrm{fs}$, respectively. To quantify the polarization contrast under each condition, the signal beam is split by a polarizer and detected by a pair of SiPM detectors, which resolve the horizontally and vertically polarized photon numbers (see Figure~\ref{figure2}). While communication typically requires the accumulation of tens of photons per bit to establish reliable statistics, the method retains a key aspect of quantum-level security, as decoy-state or alternate protocols remain viable at fluences of $\sim$1~photon per pulse \cite{scaraniQuantumCryptographyProtocols2004,hoi-kwongloSecurityQuantumKey2007,yinSecurityQuantumKey2016,diamantiPracticalChallengesQuantum2016,mizutaniQuantumKeyDistribution2019} and with imperfect detectors \cite{curtyEffectFiniteDetector2004,d.gottesmanSecurityQuantumKey2004,zhangSecurityProofPractical2021}. As a proof of concept, we examine how spectral shifts and polarization effects associated with biexciton formation can be exploited to minimize the number of photons required to securely transmit information.

Signals associated with the higher-energy biexciton resonance are selectively measured using 500-nm bandpass filters, where the number of detected photons is resolved using SiPM detectors. As shown in Figure~\ref{figure8}(a), the oscilloscope traces exhibit discrete voltage peaks corresponding to integer multiples of single photoelectron events. To quantify the detected signal, the total numbers of horizontally and vertically polarized photons accumulated over $M$ pulses are computed as
\begin{equation}
N_{H}^{(M)} = \sum_{k=1}^{M} n_{k,H}\!\left(T, \lambda_{\mathrm{det}}, \theta_{\mathrm{qwp}}\right)
\end{equation}
and
\begin{equation}
N_{V}^{(M)} = \sum_{k=1}^{M} n_{k,V}\!\left(T, \lambda_{\mathrm{det}}, \theta_{\mathrm{qwp}}\right).
\end{equation}
The polarization contrasts plotted in Figures~\ref{figure8}(b) and \ref{figure8}(c) reflects the imbalance between horizontal and vertical components:
\begin{equation}
\begin{split}
&P^{(M)}\!\left(T, \lambda_{\mathrm{det}}, \theta_{\mathrm{qwp}}\right) \\
&= \frac{N_{H}^{(M)}\!\left(T, \lambda_{\mathrm{det}}, \theta_{\mathrm{qwp}}\right)
- N_{V}^{(M)}\!\left(T, \lambda_{\mathrm{det}}, \theta_{\mathrm{qwp}}\right)}
{N_{H}^{(M)}\!\left(T, \lambda_{\mathrm{det}}, \theta_{\mathrm{qwp}}\right)
+ N_{V}^{(M)}\!\left(T, \lambda_{\mathrm{det}}, \theta_{\mathrm{qwp}}\right)}.
\end{split}
\end{equation}
Although the magnitude of $P^{(M)}\!\left(T, \lambda_{\mathrm{det}}, \theta_{\mathrm{qwp}}\right)$ varies slightly from day to day due to minor alignment differences between SiPM detectors, the differences in polarization contrasts measured at delay times $T = 0$ and 500~fs (red and blue lines in Figures~\ref{figure8}(b) and \ref{figure8}(c)) are highly reproducible. 

\begin{figure*}[!t]
\centering
\includegraphics[width=0.85\textwidth]{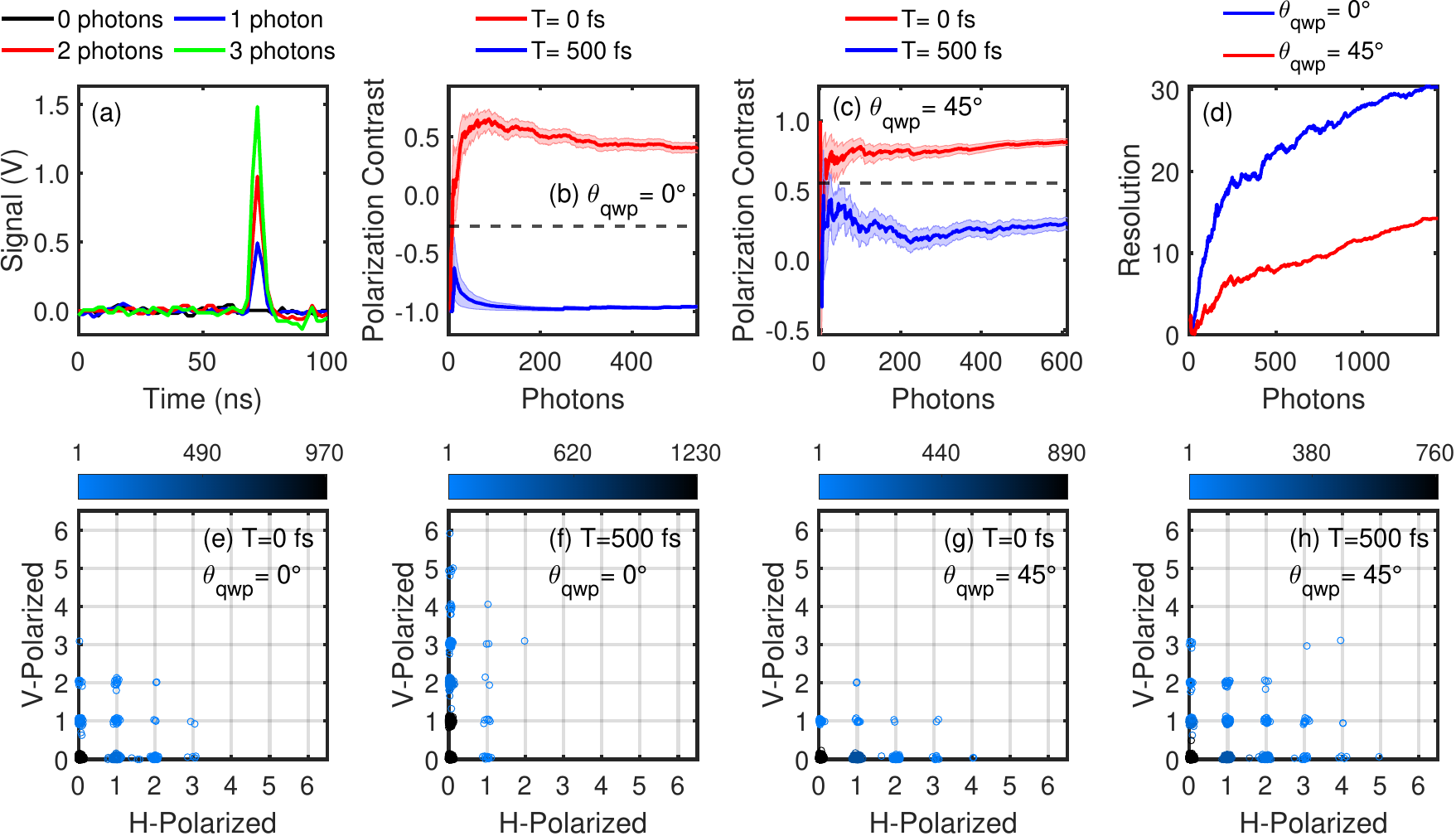}
\caption{Photon-resolved transient grating signals measured at a detection wavelength of 540~nm. (a) Signals acquired with the silicon photomultiplier detectors represent integer multiples of the voltage corresponding to a single photoelectron. (b)–(c) Convergence of the mean values $\bar{P}^{(M)}$ and standard errors $\sigma_{P}^{(M)}$ for polarization contrasts measured at the indicated pump–probe delay times and quarter-wave plate angles $\theta_{\mathrm{qwp}}$. (d) Resolution of the polarization contrasts $R_\mathrm{res}^{(M)}$ acquired for quarter-wave plate angles of $\theta_{\mathrm{qwp}} = 0^{\circ}$ and $45^{\circ}$. (e)–(h) Total numbers of horizontally and vertically polarized photons accumulated at the indicated pump–probe delay times and quarter-wave plate angles.}
\label{figure9}
\end{figure*}

To assess the reliability and efficiency of quantum state discrimination in our implementation, it is essential to quantify the uncertainty in the measured polarization contrasts. The standard errors for the polarization contrasts shown in Figures~\ref{figure8}(b) and \ref{figure8}(c) are given by
\begin{equation}
\begin{split}
&\sigma_{P}^{(M)}\!\left(T, \lambda_{\mathrm{det}}, \theta_{\mathrm{qwp}}\right)
=    \sqrt{\frac{1}{M\left(M-1\right)}} \\
&\quad\times
   \sqrt{
      \sum_{k=1}^{M}
      \left[
         P_{k}\!\left(T, \lambda_{\mathrm{det}}, \theta_{\mathrm{qwp}}\right)
         - \bar{P}^{(M)}\!\left(T, \lambda_{\mathrm{det}}, \theta_{\mathrm{qwp}}\right)
      \right]^{2}
   }.
\end{split}
\label{eq:sigmaP}
\end{equation}
where the polarization contrast for signal pulse $k$ is
\begin{equation}
\begin{split}
&P_{k}\!\left(T, \lambda_{\mathrm{det}}, \theta_{\mathrm{qwp}}\right) \\
&= \frac{n_{k,H}\!\left(T, \lambda_{\mathrm{det}}, \theta_{\mathrm{qwp}}\right)
- n_{k,V}\!\left(T, \lambda_{\mathrm{det}}, \theta_{\mathrm{qwp}}\right)}
{n_{k,H}\!\left(T, \lambda_{\mathrm{det}}, \theta_{\mathrm{qwp}}\right)
+ n_{k,V}\!\left(T, \lambda_{\mathrm{det}}, \theta_{\mathrm{qwp}}\right)},
\end{split}
\label{eq:Pk}
\end{equation}
and the mean contrast accumulated over $M$ pulses is
\begin{equation}
\begin{split}
&\bar{P}^{(M)}\!\left(T, \lambda_{\mathrm{det}}, \theta_{\mathrm{qwp}}\right) \\
&= \frac{1}{M}
\sum_{k=1}^{M}
\frac{n_{k,H}\!\left(T, \lambda_{\mathrm{det}}, \theta_{\mathrm{qwp}}\right)
- n_{k,V}\!\left(T, \lambda_{\mathrm{det}}, \theta_{\mathrm{qwp}}\right)}
{n_{k,H}\!\left(T, \lambda_{\mathrm{det}}, \theta_{\mathrm{qwp}}\right)
+ n_{k,V}\!\left(T, \lambda_{\mathrm{det}}, \theta_{\mathrm{qwp}}\right)}.
\end{split}
\label{eq:Pmean}
\end{equation}
The corresponding resolution plotted in Figure~\ref{figure8}(d) quantifies the distinguishability of polarization contrasts measured at the encoded delays $T_0$ and $T_1$:
\begin{equation}
\begin{split}
&R_\mathrm{res}^{(M)}\!\left(\lambda_{\mathrm{det}},\theta_{\mathrm{qwp}}\right) \\
&= \frac{\big|
\bar{P}\!\left(T_0, \lambda_{\mathrm{det}}, \theta_{\mathrm{qwp}}\right)
- \bar{P}\!\left(T_1, \lambda_{\mathrm{det}}, \theta_{\mathrm{qwp}}\right)
\big|}
{\sqrt{
\left[\sigma_{P}^{(M)}\!\left(T_0, \lambda_{\mathrm{det}}, \theta_{\mathrm{qwp}}\right)\right]^{2}
+ \left[\sigma_{P}^{(M)}\!\left(T_1, \lambda_{\mathrm{det}}, \theta_{\mathrm{qwp}}\right)\right]^{2}}}.
\end{split}
\label{eq:R}
\end{equation}
While the resolution improves with the number of signal pulses, the results presented in Figures~\ref{figure8}(b)--\ref{figure8}(d) indicate that accumulating tens of photons is sufficient to distinguish the encoded polarization states, demonstrating the feasibility of this quantum communication approach.

Quantized signal detection enables binning of the numbers of horizontally and vertically polarized photons detected per laser shot. The photon number axes in Figures~\ref{figure8}(e)--\ref{figure8}(h) are generated by dividing the peaks of the detector signals by the voltage corresponding to a single photoelectron ($\sim$0.6~V per photon). Although the average intensity corresponds to approximately one photon per pulse, individual laser shots yield between zero and five detected photons, with zero-photon events being the most probable. While the two detectors register coincident events, the arrivals of horizontally and vertically polarized photons are statistically uncorrelated. Additionally, second-order autocorrelation values computed from the photon number distributions yield slightly bunched statistics, with $g^{(2)} = 1.0$--$1.5$. This behavior is tentatively attributed to the nonlinear filamentation and supercontinuum generation processes used to produce the signal pulses \cite{erkintaloStatisticalInterpretationOptical2010,lushnikovNonGaussianStatisticsMultiple2010}. Because the signal processing algorithm sums over all photon number states, a more detailed quantum optical analysis is not required for the present application.

Photon-resolved transient grating signals measured with 540-nm bandpass filters exhibit signatures of the short-lived red-shift in the RRVH tensor element discussed above. In Figure~\ref{figure9}, polarization contrasts are compared at delay times $T = 0$ and 500~fs for quarter-wave plate angles $\theta_{\mathrm{qwp}} = 0^{\circ}$ and $45^{\circ}$. As anticipated, the values of $P^{(M)}\!\left(T, \lambda_{\mathrm{det}}, \theta_{\mathrm{qwp}}\right)$ measured at $\theta_{\mathrm{qwp}} = 0^{\circ}$ differ by approximately 1.3 between the two delay times (see Figure~\ref{figure9}(b)). This enhancement in the polarization contrast at $\theta_{\mathrm{qwp}} = 0^{\circ}$ arises from a transient increase in the horizontally polarized component of the signal field near 540~nm at $T = 0$~fs (see Figure~\ref{figure7}). Consequently, the polarization contrast reverses sign when the signal field becomes vertically polarized at $T = 500$~fs due to spin relaxation. The resolutions plotted in Figure~\ref{figure9}(d) underscore the advantages of this nonlinear optical effect for quantum information transmission at 540~nm with $\theta_{\mathrm{qwp}} = 0^{\circ}$. Assuming equal standard errors, the difference in the mean polarization contrasts observed in Figure~\ref{figure9}(b) suggests that quantum bits can be transmitted using significantly fewer photons compared to the conditions presented in Figures~\ref{figure8}(b)--\ref{figure8}(c) and \ref{figure9}(c).

\begin{figure*}[!t]
\centering
 \includegraphics[width=0.7\linewidth]{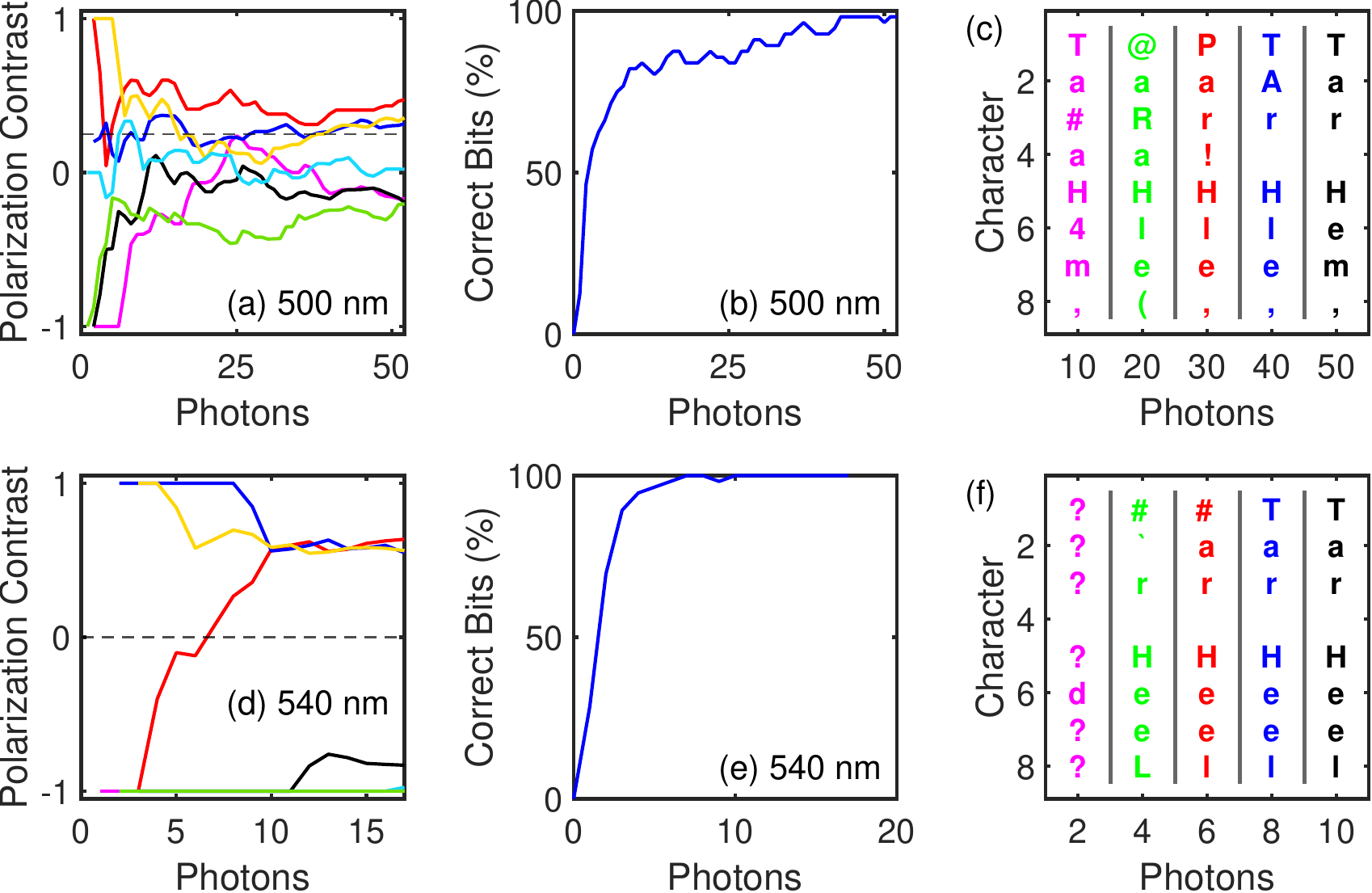}
\caption{The eight ASCII characters spelling the official nickname for UNC athletic teams, ``Tar Heel,'' are transmitted using four-wave mixing signal photons generated by (PEA)$_2$PbI$_4$. Experimental data acquired with 500 and 540~nm signal detection are displayed in the top and bottom rows, respectively. Panels (a) and (d): Polarization contrast trajectories corresponding to the seven bits encoding the ASCII character ``T.'' Panels (b) and (e): Percentages of correctly decoded bits across the full 56-bit ASCII message. The overall message converges with approximately ten times fewer photons per bit when using 540~nm light emitted by the lower-energy biexciton resonance. Panels (c) and (f): Snapshots of the decoded ASCII message as a function of the number of accumulated photons per bit.}
\label{figure10}
\end{figure*}

The quantum communication protocol is implemented by subdividing the total number of signal pulses into bits. To illustrate the process, the official nickname for UNC athletic teams “Tar Heel” is communicated using eight ASCII characters comprising a total of 56~bits, as shown in Figure~\ref{figure10}. Each character is represented as a seven-bit binary sequence, such as 1010100 for the capital letter ``T.'' Bit values are assigned based on polarization contrasts: values below and above the overall mean at each delay time are interpreted as~0 and~1, respectively. Data are acquired by cycling over the 56-bit sequence 800--1600~times while randomly varying the two delay times and two quarter-wave plate angles for each signal pulse. As a result, only 25\% of the signal pulses are useful for encoding the key, while the remaining 75\% are discarded during key extraction. For instance, at a detection wavelength of 500~nm, signals corresponding to the four measurement conditions shown in Figures~\ref{figure8}(b) and \ref{figure8}(c) are recorded over 800--1600~cycles of 56~signal pulses. In this scenario, bit values are determined by comparing polarization contrasts to the mean values obtained with $\theta_{\mathrm{qwp}} = 45^{\circ}$: contrasts below the mean are assigned a~0, and those above are assigned a~1. Signals measured with the non-matching quarter-wave plate angle or delay time are discarded, as they do not contribute to the transmitted key.

Trajectories of the polarization contrasts corresponding to the seven-bit ASCII code for the capital letter ``T'' are shown in Figures~\ref{figure10}(a) and \ref{figure10}(d) at detection wavelengths of 500 and 540~nm, respectively. In both cases, the greater polarization contrast measured at $T = 0$~fs is assigned to a binary~1, whereas photons acquired at $T = 500$~fs delay times are assigned to~0. Resolution of these binary values requires fewer photons when detection is performed at 540~nm, consistent with the differences observed between Figures~\ref{figure8}(d) and \ref{figure9}(d). Applying this procedure to all 56~bits enables transmission of the full eight-character ASCII message ``Tar Heel.'' The percentage of correctly assigned bits is plotted as a function of the number of accumulated photons in Figures~\ref{figure10}(b) and \ref{figure10}(e). This comparison shows that successful communication of the eight-character message is achieved with greater than 50~photons per bit at 500~nm and only 10~photons per bit at 540~nm. The improved efficiency at 540~nm arises from a distinctive nonlinear optical effect associated with the lower-energy biexciton.

\section{Concluding Remarks}

The quantum communication experiment demonstrated in this work leverages features of four-wave mixing signals, previously studied in the context of spin-flip transitions \cite{fengNonlinearOpticalSignatures2025,ganMotionalNarrowingSpin2025}, as a method for encoding information. Although relaxation of the macroscopic spin polarization influences the ellipticity of signal fields produced by all resonances, the effect is substantially amplified by interference between terms in the response function near the spectral region of the lower-energy biexciton in (PEA)$_{2}$PbI$_{4}$ and other 2D-OIHPs. We find that this aspect of the nonlinear optical response can be exploited to transmit one binary bit using approximately 10~photons, compared to more than 50~photons required at other detection wavelengths. This protocol highlights how the unique spin-dependent optical properties of 2D-OIHPs can be utilized for quantum information applications. While conventional 2D perovskite light sources such as light-emitting diodes rely on photoluminescence, which typically produces thermalized, incoherent emission unsuitable for deterministic quantum state preparation, the generation of four-wave mixing signal beams provides a practical alternative, with the added advantage of resonance enhancement via biexciton states.

Building on this demonstration, it is important to note that our light source involves weak coherent pulses, not a deterministic single-photon emitter. Nonetheless, protocols like traditional BB84 \cite{bennettQuantumCryptographyPublic2014} and SARG04 \cite{scaraniQuantumCryptographyProtocols2004} remain secure with attenuated laser sources (i.e., weak coherent states) \cite{scaraniQuantumCryptographyProtocols2004,d.gottesmanSecurityQuantumKey2004,hoi-kwongloSecurityQuantumKey2007,mizutaniQuantumKeyDistribution2019} provided that appropriate measures like phase randomization \cite{hoi-kwongloSecurityQuantumKey2007,naharImperfectPhaseRandomization2023} or decoy-state \cite{hwangQuantumKeyDistribution2003,loDecoyStateQuantum2005,wangBeatingPhotonNumberSplittingAttack2005} techniques are employed. Using four-wave mixing in 2D perovskites to generate these coherent pulses, we rely on the material’s Hamiltonian (i.e., exciton and biexciton dynamics) to imprint information directly onto the phase, polarization, and timing of the emitted photons. Rather than manipulating the signal pulses directly, this approach suggests a pathway toward secure quantum communication based on spin-dependent nonlinear optical processes in 2D-OIHPs, where the quantum information is not imposed externally but rather emerges from the internal dynamics of the material itself. Such a material-defined encoding mechanism could simplify transmitter design and allow for resonance-enhanced coherent sources to serve as compact, spin-driven photon emitters.

Although this experimental demonstration focuses on a proof of concept, our method supports rigorous quantum security analysis based on density matrices for the light source and simulated channel conditions. By reconstructing polarization-resolved signal fields from attenuated four-wave mixing measurements, it is possible to quantitatively evaluate quantum channel characteristics using metrics such as the Holevo bound \cite{a.s.holevoCapacityQuantumChannel1998}, Eve’s optimal guessing probability \cite{tomamichelTightFinitekeyAnalysis2012}, and secure key rates under decoy-state protocols \cite{wangBeatingPhotonNumberSplittingAttack2005,loDecoyStateQuantum2005}. These quantities may also be derived from a more fundamental perspective using nonlinear optical response functions that incorporate the system Hamiltonian and line broadening mechanisms \cite{fengNonlinearOpticalSignatures2025,ganMotionalNarrowingSpin2025}. While the parameters explored here were chosen based on practical considerations of polarization resolution, the broader space of pump–probe delay times and detection wavelengths offers a practical alternative for optimizing resistance to photon-splitting and related attacks. From an instrumentation standpoint, although this implementation relies on a titanium–sapphire laser system, the approach is fully compatible with room-temperature, free-space operation using compact Yb-doped femtosecond lasers operating at repetition rates in the hundreds of kilohertz.

\begin{acknowledgments}
This work is supported by the National Science Foundation under Grant Nos.~CHE-2247159 (S.F., Z.G., and A.M.) and CHE-2154791 (C.G. and W.Y.). This work was performed in part at the Chapel Hill Analytical and Nanofabrication Laboratory (CHANL), a member of the North Carolina Research Triangle Nanotechnology Network (RTNN), which is supported by the National Science Foundation under Grant No.~ECCS-2025064, as part of the National Nanotechnology Coordinated Infrastructure (NNCI).
\end{acknowledgments}

\bibliographystyle{aipnum4-1}
\bibliography{ZoteroLibraryAbbrev}

\end{document}